\documentclass{article}

\usepackage{arxiv}

\usepackage[utf8]{inputenc} 
\usepackage[T1]{fontenc}    
\usepackage{hyperref}       
\usepackage{url}            
\usepackage{booktabs}       
\usepackage{amsfonts}       
\usepackage{nicefrac}       
\usepackage{microtype}      

\usepackage{graphicx}
\usepackage{natbib}
\usepackage{doi}

\usepackage{algorithm}
\usepackage{algorithmic}
\usepackage{caption}

\title{Optimal wireless Rate and Power control in the Presence of Jammers Using Reinforcement Learning}

\author{ {\hspace{1mm}Fadlullah Raji}\thanks{Fadlullah Raji and Lei Miao.} \\
	Department of Computer Science\\
	University of South Florida \\
	Tampa, Florida, USA \\
	\texttt{fraji@usf.edu} \\
	\And
	{\hspace{1mm}Lei Miao} \\
	Department of Engineering Technology\\
	Middle Tennessee State University\\
	Murfreesboro, Tennessee, USA \\
	\texttt{lei.miao@mtsu.edu} \\
}

\date{}

\renewcommand{\headeright}{}
\renewcommand{\undertitle}{}
\renewcommand{\shorttitle}{}

\begin{document}
\maketitle

\begin{abstract}
	Future wireless networks require high throughput and energy efficiency. This paper studies using Reinforcement Learning (RL) to do transmission rate and power control for maximizing a joint reward function consisting of both throughput and energy consumption. We design the system state to include factors that reflect packet queue length, interference from other nodes, quality of the wireless channel, battery status, etc. The reward function is normalized and does not involve unit conversion. It can be used to train three different types of agents: throughput-critical, energy-critical, and throughput and energy balanced. Using the NS-3 network simulation software, we implement and train these agents in an 802.11ac network with the presence of a jammer. We then test the agents with two jamming nodes interfering with the packets received at the receiver. We compare the performance of our RL optimal policies with the popular Minstrel rate adaptation algorithm: our approach can achieve \emph{(i)} higher throughput when using the throughput-critical reward function; \emph{(ii)} lower energy consumption when using the energy-critical reward function; and \emph{(iii)} higher throughput and slightly higher energy when using the throughput and energy balanced reward function. Although our discussion is focused on 802.11ac networks, our method is readily applicable to other types of wireless networks.
\end{abstract}


\section{Introduction}
Future communication networks need to provide high data rates to users in an energy efficient way. Wi-Fi is a very popular type of wireless network, and there were 22.2 billion Wi-Fi devices in 2021 (\cite{wifides}). Therefore, a slight future improvement on Wi-Fi can significantly improve productivity and make a positive impact to the environment. IEEE 802.11, the protocol that enables Wi-Fi,  defines physical layers that can transmit data at a variety of rates. Various channel access techniques, such as Orthogonal Frequency Division Multiplexing (OFDM) or Direct Sequence Spread Spectrum (DSSS), and modulation schemes, such as Binary Phase Shift Keying (BPSK) or variants of Quadrature Amplitude Modulation (QAM), may be used at different rates. Because effects like multipath fading, shadowing, signal attenuation, and interference from other radio sources are tolerated differently by each of these, using the fastest rate regardless of the channel circumstances is not the optimal solution. 

For this reason, various rate control algorithms (either proprietary or open-source ones) that dynamically adjust the transmission rate in response to changing channel circumstances have been designed to improve the performance of wireless networks. In particular, these rate control algorithms (\cite{b1, b2, b3, b4, b5}) are primarily designed to identify the best rate and modulation scheme that yield the highest throughput. Because reliable data transmission rates and interference levels are fundamentally connected in wireless networks, transmission power control (\cite{b6, b7}) has also been used to reduce undesirable interference and to conserve energy for wireless devices, especially the battery-powered ones. Joint transmission rate and power control has been explored to take into account the trade-off between the throughput and the energy consumed (\cite{b8}). In principle, decreasing the power or raising the carrier sense threshold may help to enhance spatial reuse. By differentiating congestion from interference losses, (\cite{b9}) proposes a hybrid transmit power and carrier sense adaptation approach. When the interference occurs before the data signal, this work shows that fine-tuning the carrier detection threshold may completely remove interference-related losses. In addition, power control avoids data signal loss due to interference that occurs when the data signal is sent. (\cite{b10}) explores the trade-off between energy and latency and uses a real-time controller for the dynamic regulation of task delivery in order to minimize energy consumption while meeting a deadline for each individual task. In particular, the authors make use of the generalized critical task decomposition algorithm to identify critical tasks on an optimal sample path.

In addition to power and carrier sense management, when rate control is taken into consideration, it introduces a trade-off between spatial reuse and the transmission rate that can be sustained (\cite{b11}). A new idea, spatial back-off, was introduced in (\cite{b12}), which allows for dynamic tweaking of the carrier sensing threshold in conjunction with the Auto-Rate Fallback (ARF) algorithm in order to achieve high throughput. In particular, ARF shifts to a lower transmission rate if the measured losses exceed a threshold, then switches to a higher transmission rate after a specified number of consecutive frames are successfully sent. According to (\cite{b13}), when dealing with discrete data rates and when there are a sufficient number of power levels, controlling the power gives several benefits over carrier sensing control as compared to a continuous data rate. According to the authors, power and rate control is a technique that regulates the power and rate of a transmitter depending on the perceived degree of interference at the receiving end. It is necessary for the receiver to return this information to the transmitter, which may be accomplished by IEEE 802.11k (\cite{b14}), but is not currently supported by any of the device driver versions.

An adaptive rate and power control technique that is consistent with IEEE 802.11 operations is proposed in (\cite{b15}) where Acknowledgments (ACKs) received from the receiver are used to communicate the optimization of the transmission speed, which continues to operate utilizing two basic adaptive strategies: the maximum possible rate is assisted with the least potential power; and the lowest possible power is chosen first, then the highest rate conceivable at this power is chosen. In a related manner, Power-controlled Auto Rate Fallback (PARF) and Power-Enabled Rate Fallback (PERF) were suggested in (\cite{b16}), in which the authors extend ARF and Estimated Rate Fallback (ERF) to work with transmission power control. It is important to note that ERF is the SNR-based variant of ARF, in which each packet carries the power level, the path loss, and noise estimate from the previous packet that has been received. ERF senders estimate the SNR based on this information and establish the highest transmission rate compatible with the estimated SNR. The authors of (\cite{b16}) discovered that PARF did not work effectively when the receiver reduced the power used for ACK messages, as they predicted. In essence, this resulted in inaccurate power reduction choices at the transmitter when these ACK packets were not received. They get more reliable performance using PERF, making power and rate choices based on the SNR values. These findings are consistent with (\cite{b17}), which demonstrates that SNR-based treatments are more resilient when compared to loss-based protocols (\cite{b14}). Despite this, they conclude that in order to achieve such resilience, SNR-based methods necessitate real-time training.
\section{Related Work}
Reinforcement Learning (RL) has been researched actively for the control of transmission power and the data rates for 802.11 standards. (\cite{b18}) presented a power allocation technique based on multi-agent reinforcement learning. The paper reduces the loss function through stochastic gradient descent using a Deep Q-Network with many agents learning in parallel. The state description for each agent is the previous transmit power which describes agent $i$'s  potential contribution to the network as well as the interfering neighbors'  contributions to the network based on observations from a set of $n$ transmitters with an SNR greater than a predefined threshold and a receiver with an SNR greater than the threshold. The actions are described as discretized steps of power within a specified power range shared by all actors (i.e., all agents have the same action space). The reward function is intended to reflect each agent's direct interference contribution to the network and its penalty for interfering with all other agents, defined and interpreted as how action of agent $i$ through time slot $t$, i.e., $p_i^{(t)}$, affects the weighted sum-rate of its own and future interfered neighbors. 

The authors of (\cite{b19}) builds on (\cite{b18}) and uses an actor-critic algorithm to learn the optimal policy of a distributed power control. In actor-critic algorithms, two neural networks are designed to learn and update each other's weight regarding the experiences and state-action pairs encountered in each episode. Each transmitter is designed to be a learning agent in the system exploration, so the actors are a number of learning agents whose next state is conditioned upon the joint actions of all agents that existed as an actor. The critic is a single network described as the $Q_{target}$ used to update each learning agent's parameters after every episode. The presented work is a power allocation scheme for conventional wireless mobile networks that considers interference from other networks and distributes learning of the optimal policy through an actor-critic agent. 

Rate adaptation simulation using the standard 802.11g with a finite state ($<$ 100) was investigated in (\cite{b20}) using the SARSA algorithm with learned Q-values stored in a table. The states were defined to simulate a standard Robust Rate Adaptation Algorithm (RRAA) which minimizes the loss rate, $R_{loss}$, to achieve the desirable transmitting rate. This approach falls short since it considers standard Wi-Fi with a lower data rate than the latest advancement of the 802.11 such as the 802.11ac and 802.11ax, which has a lot more data rates encoded as the Modulation Coding Schemes (MCS). Another rate adaptation algorithm simulated in (\cite{b21}) represents the state's observation to the agent as Contention Window (CW) size of the CSMA/CS on an 802.11a standard, which has 8 MCS level: \{6, 9, 12, 18, 24, 36, 48, 54\}Mbps as the action space. Locally, the sender node has access to the observation. Each CSMA/CA node operating in Distributed Coordinated Function (DCF) mode chooses the random back-off time depending on the current CW size, ensuring that packets transmitted by other nodes do not overlap in the same manner. When a node transmits a packet for the first time in a typical CSMA/CS 802.11 protocol, it decides the minimum size of the CW, which is 15 in IEEE 802.11a. If the packets do not arrive at the receiver, the sender re-sends the packet at twice the CW size. This algorithm uses the CW size concept to discretize its states-action pairs stored in a Q-values table which is not enough representation of the state of the channel, as packets of bits could be dropped not only due to the interference but also due to the state of the channel.

This paper proposes a new paradigm that combines the control of transmission speed and the transmission power of the 802.11 Wi-Fi protocol using the recently developed 802.11ac standard. Specifically, we explore the use of an RL algorithm to observe the channel state and develop an optimal policy for transmitting packets of information in the presence of jamming nodes intended for deliberately disrupting the delivery of packets to the receiver. Compared with other related works in the literature, our contributions are three-fold: (i) Our state space includes multiple elements, including packet queue length, ACK from the receiver, battery level, CW, and back-off slots; (ii) We jointly control the transmission power and data rates of a transmitter under jamming and incorporate both energy and throughput into the reward function; and (iii) We show in simulation that our method outperforms the widely used Minstrel rate adaptation algorithm. It is worth noting that differently from the surveyed literature, our methodology combines two reward functions and offers a flexible approach to users for selecting which factor is more important (i.e., either to maximize the throughput of the system or to minimize the energy consumption of the device).
	
The organization of the rest of the paper is as follows: the methodology is presented in Section 3; Section 4 discusses the agents' training and testing results; the conclusion and future work are discussed in Section 5.

\section{Methodology}

\subsection{State observation}
The state is represented by a collection of characteristics derived from local measurements taken at the Transmitting node (Tx). These characteristics should give sufficient information about the transmitter's performance and the wireless channel. In particular, the state is defined as a tuple ($N_{t}$, $Cw, B_{fs}, R_p, B_l$):

\begin{itemize}
	\item $N_t$ (packet queue length): the percentage of the packet queue occupied by packets that are ready to be transmitted or re transmitted. The maximum number of packets that can be queued is 5000. The queue uses a First-In-First-Out (FIFO) policy, and it is full when the packets in the queue has reached 5000. To reduce the state space, we divide the queue length into 10 discrete levels, i.e., $N_t$ can only be multiples of 10 between 10 and 100. 
	
	\item $C_w$ (CW size): This defines a period of time in which the network is operating in contention mode. The larger the contention window, the larger the average back-off value, and the lower the likelihood of collisions. For a Best Effort (BE) packet delivery, the contention window duration doubles its current value when there is collision; the minimum contention value, $C_{w(min)}$, is 15, and the  maximum contention value, $C_{w(max)}$, is 1023. That is, there are 7 power of two values possible for the contention window between 15 and 1023. 
	
	\item $B_{fs}$ (back-off slots): This is the value returned from the back-off algorithm for collision resolution used to alert collision and re-transmission of packets when there are collisions during the transmission schedule. When a station enters the back-off state, it waits for an additional and randomly selected number of time slots (the random number is larger than 0 and less than the current CW maximum value). The total possible value of the slots in the state space is a slot range from 0 to 1023, inclusive or 1024 values. This is discretized into 128 categories by the given equation below
	
	\begin{equation}
		B_{fs} \leftarrow int(B_{f}/8),
	\end{equation}
	
	where:
	
	\begin{description}		
		\item $B_f=$ The returned collision resolution value
		\item $B_{fs}=$ Discretized value of the back-off slots
	\end{description}

	\item $R_p$ (acknowledged packet): this is an indication of whether last packet transmission has been acknowledged/successfully received. This value could be either 1 for success or 0 for failure. It is included in the state because it provides useful information about the quality of current wireless channel state to the agent. 
		
	\item $B_l$ (battery level): the percentage of the remaining battery of the transmitter. To reduce the state space, $B_l$ can only be a multiple of 10 between 10 and 100. We have this in the state so that the agent can possibly improve energy efficiency and extend the lifetime of the wireless node.
	
\end{itemize}
The size of the state space is 10 x 7 x 128 x 2 x 10 = 179,200.

\subsection{Reward function}
Generally speaking, RL algorithms face one difficult challenge: the selection of the reward function. The fact that reward functions are often hand-engineered and domain-specific is a major disadvantage of RL. Imitation Learning (IL) (\cite{b23}), which seeks to retrieve the expert policy explicitly, and Inverse Reinforcement Learning (IRL)(\cite{b24}), which offers a means to automatically obtain a cost feature from expert presentations, have received a lot of attention. However, both of these approaches have high computing costs, and the optimization problem of determining which compensation function ideally describes the expert trajectory is basically ill-posed (\cite{b25}).

Designing a reward function may be done in one of two ways: dense reward functions offer a reward in all states during exploration while sparse reward functions only provide a reward at the terminal state and no reward elsewhere. A dense reward is utilized in this application because it helps in providing feedback for each action performed so far by the agent, even if it has not yet converged to the optimal policy. The reward is intended to be the number of packets successfully transmitted and the total amount of energy consumed given the current change in transmission power and MCS. However, although the sent packets in a state may be an integer number, the energy used during a state transition - 5ms may be a value in millijoules that has a negligible effect on the agent's choice. As a result, it is necessary to normalize the throughput and energy so that the energy and packets received may be utilized to notify the agent of its choice. Our reward function is computed using the convex combination in the reward equation below. The energy consumed is multiplied by -1 to ensure that the agent prefers and considers a data rate and transmission power that uses less energy while still delivering packets at a high rate.

\begin{equation}
	Reward = \lambda \cdot \frac{  R_p \times 100}{ N_T} +  (1 - \lambda) \cdot \frac{ -1 \times E_c \times 100}{E_T}
	\label{eq:reward}
\end{equation}

where:

\begin{description}
	\item $R_p=$ Number of received packet in a state
	\item $E_c=$ Energy consumed in a state
	\item $N_T=$ Total number of packets available for transmission
	\item $E_T=$ Total energy available at the transmitter
	\item $\lambda=$ Weighted multiplier
	
\end{description}

When $\lambda$ is set to a value higher than 0.5, the agent gives priority to throughput over the consumed energy; when $\lambda$ is set to a value less than 0.5, the agent gives priority to the consumed energy over throughput. 

The total throughput during simulation is calculated as
\begin{equation}
	\label{eq:throughput}
	Throughput (Mbps) = \frac{N_p \times P_s \times 8}{T_s \times 10^6},
\end{equation}

where $N_p$ is the total number of packets received during simulation, $P_s$ is the carrying payload size of a packet and $T_s$ is the total simulation time.

\subsection{Actions}
\begin{figure}[htbp]
	\centering
	\includegraphics[scale=0.3]{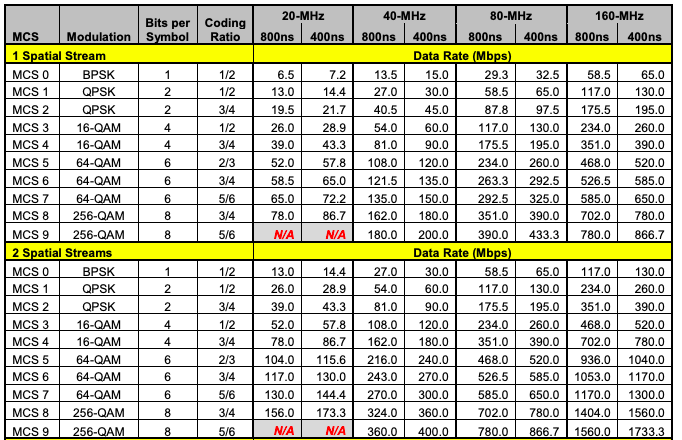}
	\caption[802.11ac PHY data rates for 1 and 2 spatial streams]{802.11ac PHY data rates for 1 and 2 spatial streams (\cite{b26})}
	\label{fig:802.11ac}
\end{figure}

The primary action of the RL agent is choosing the data rates and the transmission power. The data rates in the 802.11ac Wi-Fi protocol, also known as the Very High Throughput (VHT) rates, are represented as MCS values between 0 and 9: for one spatial stream, the minimum throughput is 58.5 Mbps at MCS level 0 and the maximum value is 780 Mbps for MCS level 9. See the whole list in Fig. \ref{fig:802.11ac}.

The transmission power has 10 levels, ranging from 1 dBm to 10 dBm. We limited the maximum power to 10 dBm since we are simulating a battery-powered device.

\subsection{The terminal state}
The study of transmission speed and power control in wireless networks is considered to be a sequential decision-making problem, since we will continue to send packets whenever they are ready for transmission. Today's wireless devices, on the other hand, mostly rely on batteries. Thus, we make the RL system an episodic problem with a terminal state, a condition that ends the training episode. In order to model the scenario of battery depletion and let the agent learn how to use the energy efficiently, we utilize the battery level as a primary termination state: an episode terminates if the battery level falls below 10 percent. Additionally, we are concerned with dropped packets, which are not part of the reward function. Therefore, we have a secondary terminal state that ends training if the number of dropped packets increases over a predefined proportion: 5\%s of the total packets arriving at the packet queue. This secondary terminal state acts like a penalty: the agent needs to choose actions that yield less dropped packets in order to maximize the reward.

\subsection{Learning agent}
Let us denote the action space by A and the state space by S. A total of 100 actions is given in a flattened vector (10 x 10) of the MCS value and the transmission power. Let $ s_i \in S$ represent the state of the system at a particular time step t. In this condition, the agent will perform the action $a_i \in A$ that has the highest Q-value, i.e., $a_i=argmax_{a \in A} Q(s_i, a)$. The Q values of all state-action pairs are maintained in a so-called look-up table, which is updated on a regular basis in accordance with the reward function.

\begin{table}[h]
	\centering
	\caption{Storage complexity}\label{tab:tab1} 
	\label{table:q_complexity}
	\begin{small}
		\begin{tabular}{ll}
			\hline
			Set / Function & Cardinality \\ 
			& \\
			\hline
		
			State Space:
			$  N_t  \times C_w \times B_{fs} \times R_p\times B_l $  & $10 \times 7 \times 128 \times 2 \times 10 $   \\
		
			Action Space: $ A = P_w \times MCS $ & $10 \times 10   $       \\ 
			Q-value function: $Q^t(s, a)$ & $17,920,000 $  \\
			\hline
			
		\end{tabular}
	\end{small}
\end{table}

We use SARSA, an on-policy RL algorithm that updates the Q-function using the experienced trajectory $(s_t,a_t,r_t,s_{t+1},a_{t+1}, r_{t+1}, \dots$). The implementation of our algorithm uses a SARSA update rule for the Q-value computation of each state $S$ in the look up table. The SARSA algorithm is considered so that the agent is informed about how an action can create an interference or can cause a packet to be dropped in the next state. Mathematically, the update rule of the Q-table using the SARSA algorithm is given as

\begin{equation}
    	Q^{new}_{(s_i, a_i)} = (1 - \alpha) \times Q(s_i, a_i) +  
		\alpha ( r(s_i, a_i) + \gamma^{\Delta t}\cdot Q(s_{i + 1}, a_{i+1}) - Q(s_i, a_i))
		\label{eq:sarsn}
\end{equation}

The storage complexity calculation can be found in Table \ref{table:q_complexity}, and the training algorithm is described in Algorithm \ref{alg:alg2}. Exploration and exploitation are the two strategies in RL. In exploration, an agent seeks additional knowledge about the environment to verify the presence of better decisions; in exploitation, the agent adopts the best actions known so far to maximize the reward function. More exploration implies the agent is more likely to choose many sub-optimal actions, lowering its performance. However, the agent may be completely unaware of alternative actions that result in better long-term results if the agent only performs exploitation. $\epsilon$ is a probability of selecting random actions and is initially set to 1 so that the agents can navigate and iterate through the actions randomly. $\epsilon^{'}$ is a multiplier reducing the probability of selecting random actions over time. 

\begin{algorithm}
	\caption{The SARSA Algorithm}
	\label{alg:alg2}
	\begin{algorithmic}[1]
		\STATE Initialize Simulation Environment 
		\STATE  Initialize Q-Table and Agent Methods
		\STATE Initialize Number of Episodes 
		\STATE Initialize Epsilon

		\FOR{ iteration = 1, 2 ..., Episodes}
		\STATE $TotalReward \gets 0$
		\WHILE{Transmission is not completed}
		\STATE $Action \gets Get \quad Action (A_t)  $
		\STATE Perform Action in Simulation Environment
		\STATE Collect Next State, Reward, Done
		\STATE $Action_{(t+1)} \gets Get \quad Next State Action (A_{t+1}) $
		\STATE Update Q-Value using \ref{eq:sarsn} and the Q-Table
		\STATE $Epsilon \quad \epsilon \gets \epsilon *\epsilon'$ 
		\STATE $TotalReward + = Reward$
		\ENDWHILE
		\ENDFOR
	\end{algorithmic}
\end{algorithm}

\section{RESULTS AND DISCUSSION}
We use RL to optimize a joint reward function of the throughput and the transmission energy. The experiment and testing are carried out on wireless nodes using the 802.11ac standard, but our results are also applicable to other types of wireless networks. The environment where the agent learns the optimal policy is created in the well-known NS3(\cite{b22}) software, which is a free and open-source simulation tool popular for simulating discrete events of networks and network protocols. It offers a collection of models that aim to give a precise MAC-level and PHY-level implementation of the 802.11 protocols. The system environment is created using C++, and the agent is implemented using Python.

\subsection{Simulation testbed}
\begin{figure}[htbp]
	\centerline{\includegraphics[scale=0.5]{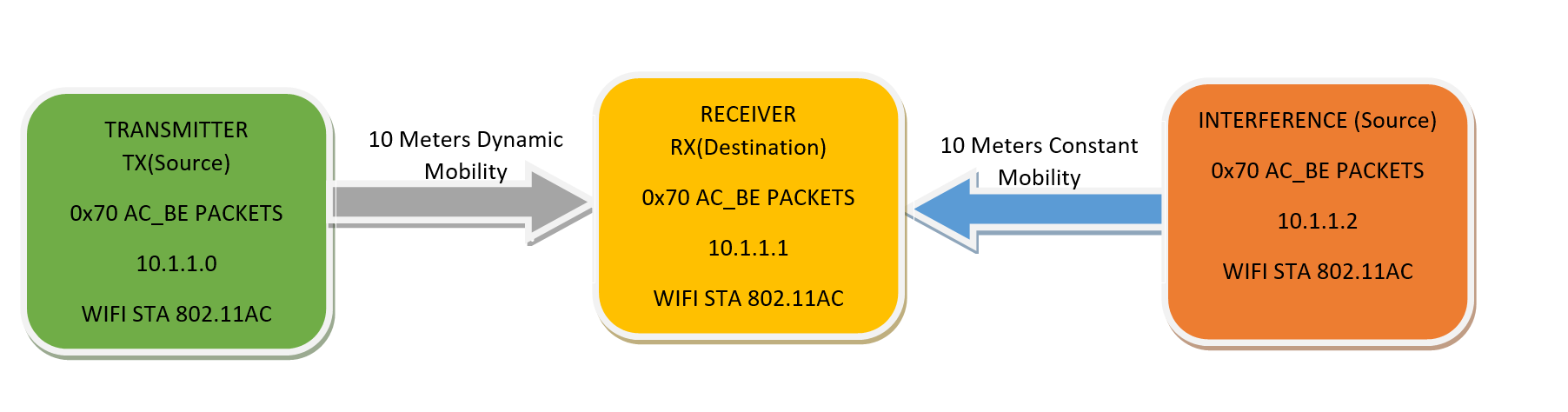}}
	\caption[Training Simulation Setup]{Training simulation setup. Transmitter (TX) continuously transmits packets to the Receiver(RX). Interference Node (IX) transmits at a random time, t, to the receiver without carrier sensing.}
	\label{fig:setup}
\end{figure}

The Transmitter (Tx) and Receiver (Rx) are the primary nodes in the simulation's testbed, and it also includes a third node and a fourth node (for testing purposes only), known as the interferers or jammers which transmit using a separate port to randomly interfere with the transmitter. See Fig. \ref{fig:setup} for the network topology. The transmitter transmits packets to the receiver, which receives these packets with a configurable Receiving Signal Strength (RSS). The interferer does not perform carrier sensing and can be considered as a hostile jamming node which randomly transmits packets in order to disrupt the transmission between the transmitter and the receiver. This interferer node is added mainly to allow the exploration of all interference states, which will aid the learning agent in exploring even more state space. The physical layer (PHY) is configured and evaluated with a spectrum utilization (i.e., the aggregated throughput of all nodes operating in the spectrum) as a function of channel width 160MHz in the 5GHz band. 

The channel is established using a simulated Friis propagation model to replicate Line-Of-Sight (LOS) path loss in a free space environment.
For the duration of the state event - 5 milliseconds, the data rate is regulated and controlled at the MAC layer, which is intended as an ad hoc Wi-Fi MAC with a constant rate Wi-Fi manager for the RL environment and may be replaced with the Minstrel algorithm or other control algorithms.

The mobility of both transmitter and the receiver, as well as the distance between them, are designed to be adjustable, with either both nodes remaining stationary throughout the simulation or the destination (Rx) dynamically moving around a desired distance to the transmitter to simulate a real-world environment where a station device or mobile hot spot may be dynamically moving away or towards the transmitter.
Best effort User Datagram Protocol (UDP) packets are generated and transmitted from the transmitter to the receiver in the simulation. We use UDP because compared with the Transmission Control Protocol (TCP), UDP is a simpler protocol that does not need connection setup delays, flow control, or re-transmission. The environment parameters is shown in Table \ref{table:q_storage}.

\begin{table}[h]
	\centering
	\caption{Simulation settings}\label{table:q_storage} 
	\begin{small}
		\begin{tabular}{ll}
			\hline
			Settings& Values\\ 
			& \\
			\hline
			\hline
			\# of nodes & 4 \\
		
			Wi-Fi Protocol & 802.11AC\\
		
			Channel & Multi Modal Spectrum Channel \\
			
			Channel width  & 160MHz \\
		
			Default Frequency& 5200 MHz \\
			
			Clear Channel Assessment (CCA) Threshold  & 82.0 dBm \\ 
			
			Guard Interval  & 800ns        \\ 
			
			Loss Model       & Friis Propagation Loss Model  \\
		
			Fading Model &      Nakagimi Loss Model    \\
		
			Mac Model  & HT-MAC        \\ 
	
			EDCA access Category  & Best Effort       \\ 
			
			Channel Delay Model   & Constant Speed Propagation Model  \\ 
			Ad hoc Mode   & Ad hoc QOS Supported  \\ 
		
			Mobility   & Random Walk and Constant Position Mobility  \\ 
			
			IP Network/Transport   & IPv4/UDP \\ 
			
			Payload Size    & 1472  \\ 
		 
			Simulation Time    & 10 Seconds  \\ 
			
			Packet Queue Size    & 5000  \\ 
		 
			Queue Maximum Delay    & 1 Second  \\ 
		\hline
		\end{tabular}
	\end{small}
\end{table}


\subsection{Energy model}
The Wi-Fi radio energy model included in the NS-3 software is used to model the network device's energy consumption, which is expressed as a percentage of the total energy consumed. The model includes all the conceivable states of the PHY layer, including idle, busy, Tx, Rx, ChannelSwitch, sleep, and off. Each of these states has a corresponding current demand value (measured in amperes). When a Wi-Fi radio energy model PHY listener is registered to the Wi-Fi PHY, it is notified of any Wi-Fi PHY state changes. When a state transition happens, the energy used in the previous state is computed, and the energy source is notified to update its remaining energy. The Wi-Fi Tx current model enables us to compute the transmit current demand as a function of the nominal transmit power (in dBm). As a result, the Wi-Fi radio energy model PHY listener receives information about the nominal Tx power needed to transmit the current frame and communicates this information to the Wi-Fi Tx current model, which is responsible for updating the current demand during transmission in the Tx state.

\begin{table}[h]
	\centering
	\caption{Agents' parameters}\label{tab:tab1} 
	\label{table:param}
	\begin{small}
		\begin{tabular}{llll}
			\hline
			
			 Learning Agents &  $\lambda$ & $\gamma$ &$\alpha$\\ 
		    \hline
		    \hline
			Throughput Optimizing & 0.8 & 0.9 &0.2 \\
		
			Energy \& Throughput Optimizing & 0.5 & 0.9 &0.2\\
		
			Energy Optimizing & 0.2 & 0.9 & 0.2 \\
			\hline
		\end{tabular}
	\end{small}
\end{table}

\subsection{Training results}

The functionality and performance of our RL algorithm are experimentally validated. As previously stated, a dictionary data structure is used to implement the Q-table in Python. The simulations are divided into two phases: training and testing. We train three agents using different values of the weighting factor $\lambda$, and the agent parameters are shown in Table \ref{table:param}. We included only one interference/jamming node in the training of each agent as shown in Fig. \ref{fig:setup}.  The training was carried out in episodes, with each episode running the simulation for  maximum 10 seconds or until the terminal state is reached, i.e., either packet loss is more than 5\% or the battery level is less than 10\%. In training, packets arrive at a constant rate of 60 000 packets per second. Each packet has 1472 bytes.

\begin{figure}
	\begin{minipage}{.5\textwidth}
	\centering
	\begin{tabular}{@{}c@{}}
		\includegraphics[width=.9\linewidth,height=150pt]{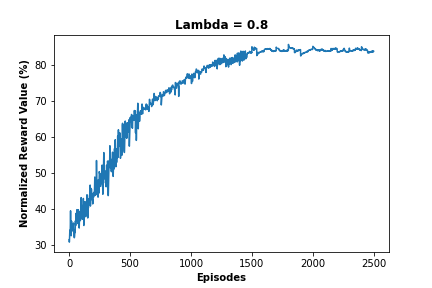} \\[\abovecaptionskip]
		\small (a) Throughput-optimizing Agent
	\end{tabular}
	
	\vspace{\floatsep}
	
	\begin{tabular}{@{}c@{}}
		\includegraphics[width=.9\linewidth,height=150pt]{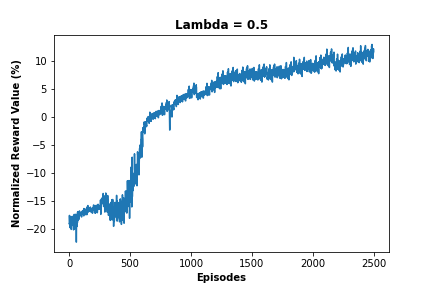} \\[\abovecaptionskip]
		\small (b) Energy and throughput optimizing agent
	\end{tabular}

	\vspace{\floatsep}
	
	\begin{tabular}{@{}c@{}}
		\includegraphics[width=.9\linewidth,height=150pt]{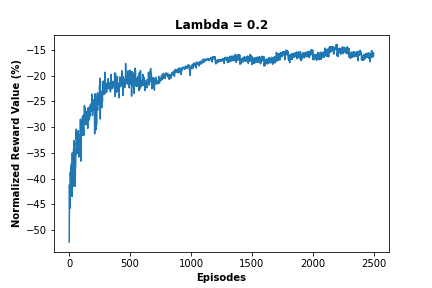} \\[\abovecaptionskip]
		\small (c) Energy-optimizing agent
	\end{tabular}
	
	\caption[SARSA Training Loop and Rewards Obtained]{SARSA rewards vs. episodes in training}\label{fig:Q-leanTraining}
	\end{minipage}%
	\begin{minipage}{.5\textwidth}
	\centering
	\begin{tabular}{@{}c@{}}
		\includegraphics[width=.9\linewidth,height=150pt]{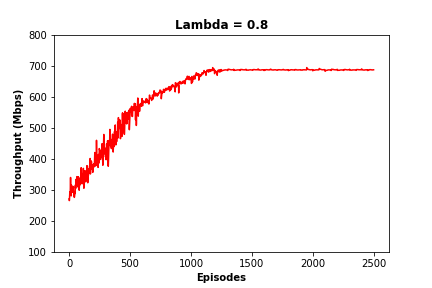} \\[\abovecaptionskip]
		\small (a) Throughput-optimizing agent
	\end{tabular}
	
	\vspace{\floatsep}
	
	\begin{tabular}{@{}c@{}}
		\includegraphics[width=.9\linewidth,height=150pt]{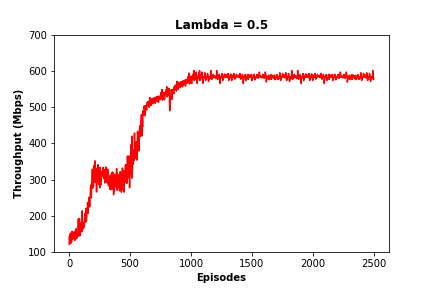} \\[\abovecaptionskip]
		\small (b) Energy and throughput optimizing agent
	\end{tabular}

	\vspace{\floatsep}
	
	\begin{tabular}{@{}c@{}}
		\includegraphics[width=.9\linewidth,height=150pt]{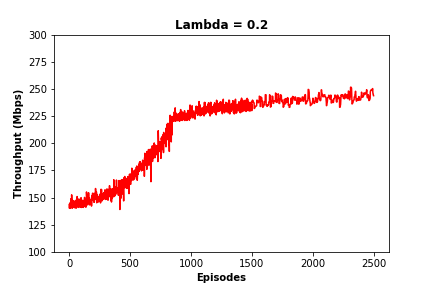} \\[\abovecaptionskip]
		\small (c) Energy-optimizing agent
	\end{tabular}
	
	\caption[Throughput vs. episodes]{Throughput vs. episodes}
	\label{fig:throughputs}
	\end{minipage}
\end{figure}

The rewards obtained in training using the SARSA algorithm is shown in Fig. \ref{fig:Q-leanTraining} where the vertical axis is the normalized reward shown in \ref{eq:reward}. The reward is low at the beginning of training (when the agent does not know the optimal policy) and then converges to the maximum over episodes after the optimal policy is learned. In Fig. \ref{fig:Q-leanTraining}(a), The weighted factor $\lambda$ is set to 0.8 to emphasize the importance of optimizing the throughput over the energy expended at the transmitter. In Fig. \ref{fig:Q-leanTraining}(b), $\lambda$ is set to 0.5, and throughput and energy consumed have equal weight. With $\lambda$ set to 0.2 in Fig. \ref{fig:Q-leanTraining}(c), energy consumption has higher priority than the throughput.

Looking closely at these figures, the reward value in Fig. \ref{fig:Q-leanTraining}(a) increases and converges to over 80 percent while the reward values in figures \ref{fig:Q-leanTraining}(b) and \ref{fig:Q-leanTraining}(c) are much lower. This is because energy consumption corresponds to a negative reward in \ref{eq:reward}. Between episodes 300 and 400 in Fig. \ref{fig:Q-leanTraining}(b), there is a significant amount of dip (i.e., the rewards goes down after the initial increase). This dip is due to the packet loss of 5\% terminal state: when 5\% packet loss occurs too soon, the agent attempts to increase the packet loss by raising the transmission rate and power; however, increasing the transmission rate and power too much makes energy consumption become worse. The agent eventually learns the best policy, resulting in an increase in the reward.

Fig. \ref{fig:throughputs} shows the increasing throughput over episodes in training as the agent learns to interact with the environment and takes actions that maximize the reward. Fig. \ref{fig:throughputs}(a) is the result from the agent with 0.8 $\lambda$ value. An examination of this figure reveals that the model learns directly to enhance throughput, as evidenced by the observation that it increases over time with far less fluctuation than figures \ref{fig:throughputs}(b) and \ref{fig:throughputs}(c). Furthermore, as shown in figures \ref{fig:throughputs}(b) and \ref{fig:throughputs}(c), the throughput increases until it converges to 600 Mbps and 250Mbps, less than 700 Mbps, the maximum throughput in Fig. \ref{fig:throughputs}(a). Since the energy optimizing agent is more concerned about the energy it uses, there are far fewer packets delivered when compared to the other agents that put more focus on optimizing the throughput.

\begin{figure}
	\begin{minipage}{.5\textwidth}
	\centering
	\begin{tabular}{@{}c@{}}
		\includegraphics[width=.9\linewidth,height=150pt]{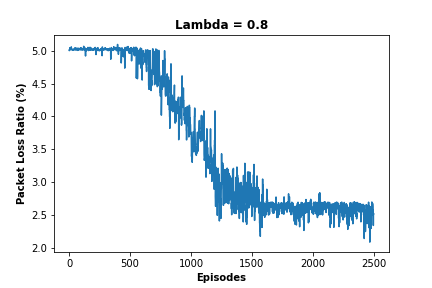} \\[\abovecaptionskip]
		\small (a) Throughput-optimizing agent
	\end{tabular}
	
	\vspace{\floatsep}
	
	\begin{tabular}{@{}c@{}}
		\includegraphics[width=.9\linewidth,height=150pt]{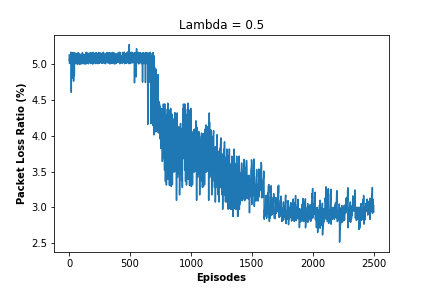} \\[\abovecaptionskip]
		\small (b) Energy and throughput optimizing agent
	\end{tabular}

	\vspace{\floatsep}
	
	\begin{tabular}{@{}c@{}}
		\includegraphics[width=.9\linewidth,height=150pt]{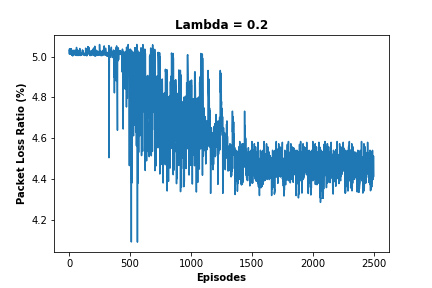} \\[\abovecaptionskip]
		\small (c) Energy-optimizing agent
	\end{tabular}
	
	\caption[Packet loss during training]{Packet loss during training}
	\label{fig:packetloss}
	\end{minipage}%
	\begin{minipage}{.5\textwidth}
	\centering
	\begin{tabular}{@{}c@{}}
		\includegraphics[width=.9\linewidth,height=150pt]{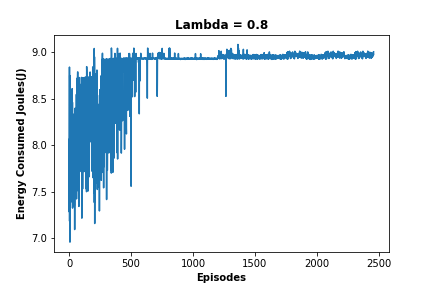} \\[\abovecaptionskip]
		\small (a) Throughput-optimizing agent
	\end{tabular}
	
	\vspace{\floatsep}
	
	\begin{tabular}{@{}c@{}}
		\includegraphics[width=.9\linewidth,height=150pt]{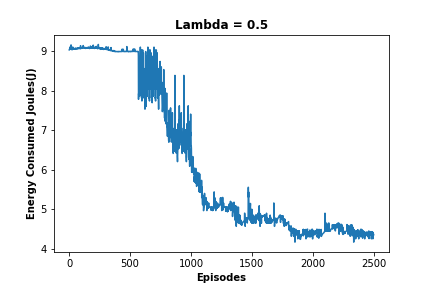} \\[\abovecaptionskip]
		\small (b) Energy and throughput optimizing agent
	\end{tabular}

	\vspace{\floatsep}
	
	\begin{tabular}{@{}c@{}}
		\includegraphics[width=.9\linewidth,height=150pt]{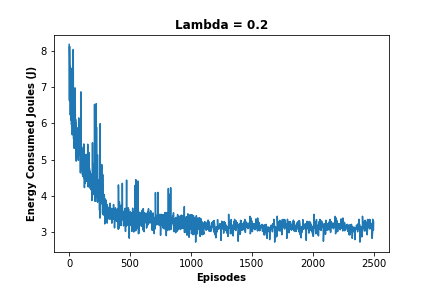} \\[\abovecaptionskip]
		\small (c) Energy-optimizing agent
	\end{tabular}
	
	\caption[Energy consumed during training]{Energy consumed during training}
	\label{fig:energyconsumed}
	\end{minipage}%
\end{figure}

The loss of packets in the MAC layer is one major factor that leads to reduced throughput. An intelligent agent should be able to reduce this loss in order to increase system efficiency and boost throughput. Fig. \ref{fig:packetloss} shows the packet loss as the number of training episodes progressively increases. In addition, the introduction of a 5\% loss rate as the the terminal state means that an episode will be ended if the loss ratio is greater than 5\% of the total packets sent, which is a severe penalty. Hence, to maximize the global reward, the agent must prioritize keeping the loss rate below 5\%.

In Fig. \ref{fig:packetloss}(a), the Throughput-optimizing agent is mostly informed about the number of packets it delivers to the receiver. Since the packet loss directly affects the packets received, the agent learned to quickly reduce this loss so that it can deliver more packets. Because the agent of Fig. \ref{fig:packetloss}(b) is concerned about the throughput and the energy consumed equally, there is more fluctuation of the packet loss until it learns to reduce the loss rate well below the terminal state threshold of 5\%. The energy-aware agent in Fig. \ref{fig:packetloss}(c) is more concerned about energy consumed, the loss rate fluctuates greatly and in general is not reduced below 4.4\%. 

The results in Fig. \ref{fig:energyconsumed} show the energy consumption on the transmitting node while training each agent. The energy-aware agent as shown in Fig. \ref{fig:energyconsumed}(c) is able to send packets with less energy consumption although the maximum achievable throughput of this agent is less than the other agents. The result of the throughput-optimizing agent shown in Fig. \ref{fig:energyconsumed}(a) consumes more energy than the other agents. Observing the energy consumption on the throughput and energy-aware agent in Fig. \ref{fig:energyconsumed}(b), we notice that the reward function is indeed informed by both throughput and the energy consumption: the agent reduces the energy consumption, compared with Fig. \ref{fig:energyconsumed}(a), and achieves higher throughput than in Fig. \ref{fig:energyconsumed}(c).

\subsection{Testing results}
In testing the optimal policy learned in training, $\epsilon$ is set to zero so that the agent only uses its trained policy to select the best action. In this phase, the Q-value computation is neither performed nor updated in the lookup table. We only match the current state and choose the action with the highest value as computed in the table.

\begin{figure}[htbp]
	\centerline{\includegraphics[scale=0.5]{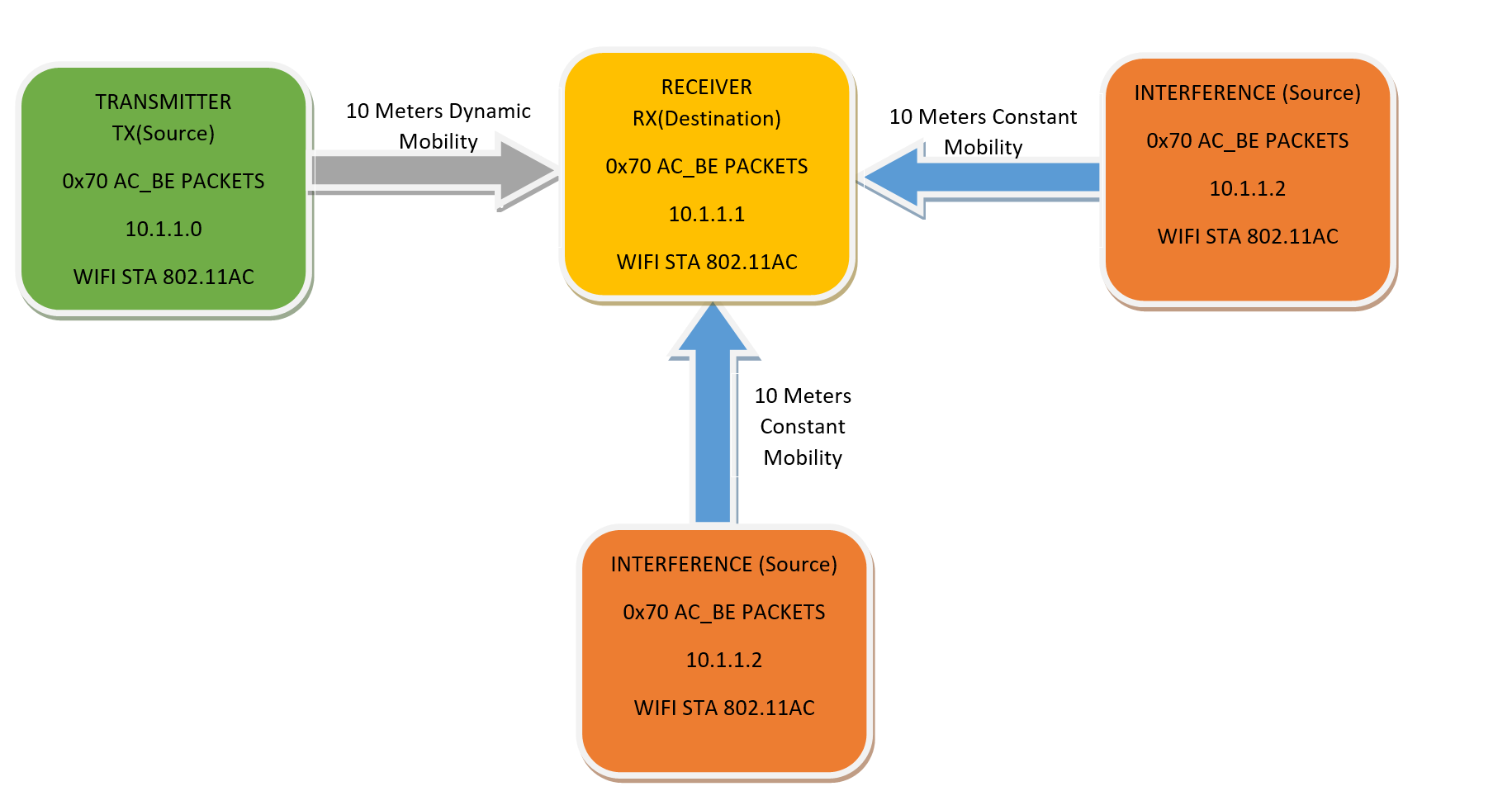}}
	\caption[Test simulation setup]{Testing simulation setup with two interference nodes. Transmitter (TX) continuously transmits packets to the Receiver (RX). Interference Nodes (IXs) transmit at a random time, t, to the receiver without carrier sensing.}
	\label{fig:2jammersetup}
\end{figure}

\begin{figure}[htbp]
	\begin{minipage}{.5\textwidth}
	\centering
	\begin{tabular}{@{}c@{}}
		\includegraphics[width=.9\linewidth,height=150pt]{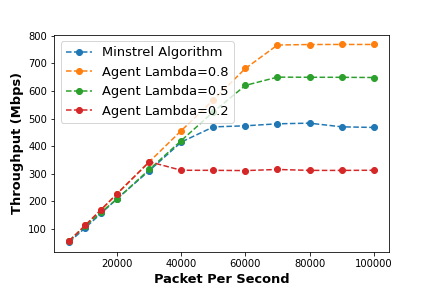} \\[\abovecaptionskip]
		\small (a) Throughput achieved
	\end{tabular}
	
	\vspace{\floatsep}
	
	\begin{tabular}{@{}c@{}}
		\includegraphics[width=.9\linewidth,height=150pt]{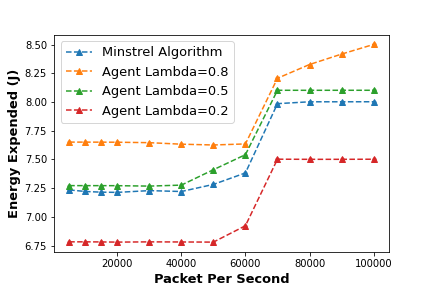} \\[\abovecaptionskip]
		\small (b) Energy consumed
	\end{tabular}
	
	\caption[Test result for 5 meter distance \\
	with one jamming node]{Test result for 5 meter distance with one \\ jamming node}
	\label{fig:test5m}
	
	\end{minipage}
	\begin{minipage}{.5\textwidth}
	\centering
	\begin{tabular}{@{}c@{}}
		\includegraphics[width=.9\linewidth,height=150pt]{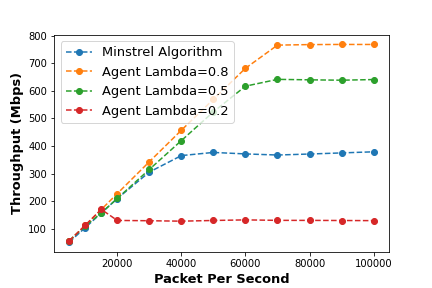} \\[\abovecaptionskip]
		\small (a) Throughput achieved
	\end{tabular}
	
	\vspace{\floatsep}
	
	\begin{tabular}{@{}c@{}}
		\includegraphics[width=.9\linewidth,height=150pt]{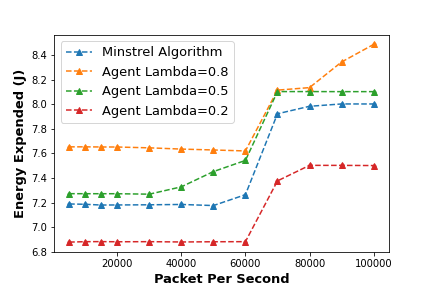} \\[\abovecaptionskip]
		\small (b) Energy consumed
	\end{tabular}
	
	\caption[Test result for 10 meter distance  \\ with one jamming node]{Test result for 10 meter distance  with one \\ jamming node}
	\label{fig:test10m}
	\end{minipage}
\end{figure}

\begin{figure}[htbp]
    
	\begin{minipage}{.5\textwidth}
	\centering
	\begin{tabular}{@{}c@{}}
		\includegraphics[width=.9\linewidth,height=150pt]{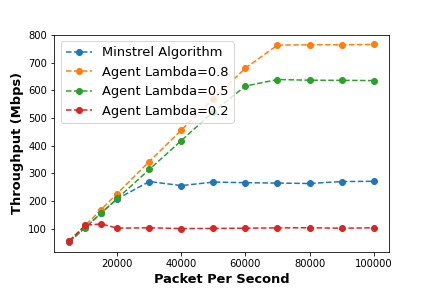} \\[\abovecaptionskip]
		\small (a) Throughput achieved
	\end{tabular}
	
	\vspace{\floatsep}
	
	\begin{tabular}{@{}c@{}}
		\includegraphics[width=.9\linewidth,height=150pt]{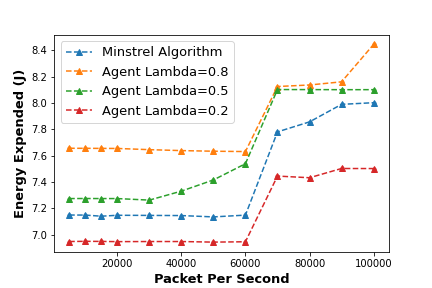} \\[\abovecaptionskip]
		\small (b) Energy consumed
	\end{tabular}
	
	\caption[Test result for 20 meter distance with one jamming node]{Test result for 20 meter distance with one \\ jamming node}
	\label{fig:test20m}
	\end{minipage}
	\begin{minipage}{.5\textwidth}
	\centering
	\begin{tabular}{@{}c@{}}
		\includegraphics[width=.9\linewidth,height=150pt]{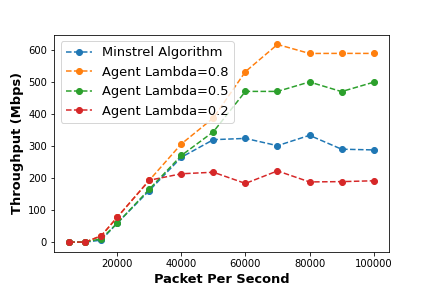} \\[\abovecaptionskip]
		\small (a) Throughput achieved
	\end{tabular}
	
	\vspace{\floatsep}
	
	\begin{tabular}{@{}c@{}}
		\includegraphics[width=.9\linewidth,height=150pt]{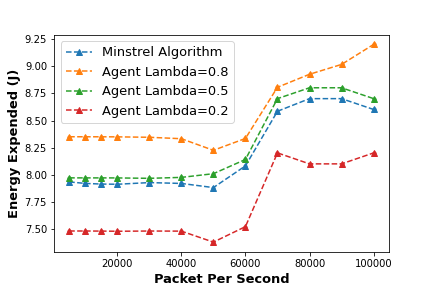} \\[\abovecaptionskip]
		\small (b) Energy Consumed
	\end{tabular}
	
	\caption[Test Result for 5 Meter Distance with two jamming nodes]{Test Result for 5 Meter Distance with two \\ jamming nodes}
	\label{fig:test5m2}
	\end{minipage}
\end{figure}

\begin{figure}[htbp]
	\begin{minipage}{.5\textwidth}
	\centering
	\begin{tabular}{@{}c@{}}
		\includegraphics[width=.9\linewidth,height=150pt]{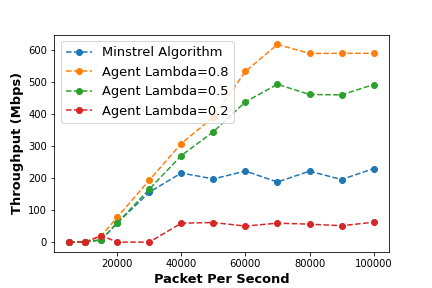} \\[\abovecaptionskip]
		\small (a) Throughput achieved
	\end{tabular}
	
	\vspace{\floatsep}
	
	\begin{tabular}{@{}c@{}}
		\includegraphics[width=.9\linewidth,height=150pt]{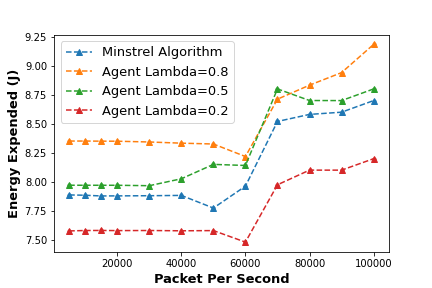} \\[\abovecaptionskip]
		\small (b) Energy consumed
	\end{tabular}
	
	\caption[Test result for 10 meter distance with two jamming nodes]{Test result for 10 meter distance with two \\ jamming nodes}
	\label{fig:test10m2}
	\end{minipage}
    \begin{minipage}{.5\textwidth}
	\centering
	\begin{tabular}{@{}c@{}}
		\includegraphics[width=.9\linewidth,height=150pt]{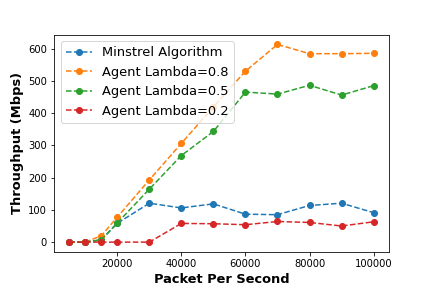} \\[\abovecaptionskip]
		\small (a) Throughput achieved
	\end{tabular}
	
	\vspace{\floatsep}
	
	\begin{tabular}{@{}c@{}}
		\includegraphics[width=.9\linewidth,height=150pt]{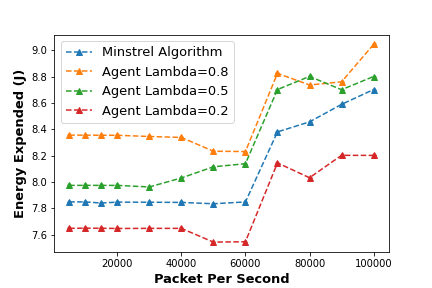} \\[\abovecaptionskip]
		\small (b) Energy consumed
	\end{tabular}
	
	\caption[Test result for 20 meter distance with two \\ jamming nodes]{Test result for 20 meter distance with two jamming nodes}
	\label{fig:test20m2}
	\end{minipage}
\end{figure}

We configured the simulation testbed to adjust the distance and mobility between the transmitter and the receiver while the jammers are set to a constant 10 meter distance to the receiver. As part of this study, we benchmark and evaluate the agents' throughput and energy consumption under one jamming node and two jamming nodes and compare them to Minstrel, a very popular rate adaptation algorithm used by many wireless drivers. In the test phase, we set up the environment with an increasing packet arrival rate, as indicated on the X-axis of figures \ref{fig:test5m}, \ref{fig:test10m}, \ref{fig:test20m}, \ref{fig:test5m2}, \ref{fig:test10m2}, and \ref{fig:test20m2}. For each arrival rate, we run 30 episodes for Minstrel and each type of training agent, but without a terminal state. In essence, each test environment remains active for precisely 10 seconds or until the battery is completely depleted. We then average the throughput and the energy expended for each arrival rate. The results for 5, 10, and 20 meters distance settings between the transmitter and the receiver and a single jamming node are shown in figures \ref{fig:test5m}, \ref{fig:test10m}, and \ref{fig:test20m}, respectively. A second jammer as shown in the network topology in Fig. \ref{fig:2jammersetup} is introduced in testing, and the learned policy and the results are presented in figures \ref{fig:test5m2}, \ref{fig:test10m2}, and \ref{fig:test20m2}.

In Fig. \ref{fig:test5m}, when the packet arrival rate is less than 30 000 packets per second, we observed that the agents attained roughly the same throughput. However, when compared to other agents, the energy-optimizing agent obtained a relatively low throughput when the packet arrival rate is 30 000 packets and higher. The Minstrel algorithm achieved close to 500 Mbps, but it is less than the throughput-optimizing and throughput and energy-optimizing agents by over 300 Mbps and 200 Mbps, respectively. In terms of energy consumption, the energy-optimizing agent is the best while Minstrel consumes less than the throughput and energy-optimizing agent, and the throughput-optimizing agent is the worst.

In Fig. \ref{fig:test10m}, we extend the distance between the transmitter and the receiver to 10 meters. Examining this figure, the energy-optimizing agent and the Minstrel algorithm achieved less throughput when compared to a distance of 5 meters. However, the throughput-aware agents with 0.8 $\lambda$ value and the throughput and energy-optimizing agent with 0.5 $\lambda$ value achieve roughly the same throughput as the case of 5 meters distance. The throughput gap between the Minstrel algorithm and the throughput-optimizing and throughput and energy-optimizing agents also widens to over 400Mbps and 300Mbps, respectively. There seems to be no significant change of energy consumption. 
 
Fig. \ref{fig:test20m} shows the result when the distance is set to 20 meters, and we observed a similar trend: the throughput gap between the Minstrel algorithm and the throughput-optimizing and throughput and energy-optimizing agents further widens to about 500Mbps and 400Mbps, respectively. 

Figures \ref{fig:test5m2}, \ref{fig:test10m2}, and \ref{fig:test20m2} are the results when we introduce a second jammer to the network topology. In Fig. \ref{fig:test5m2}, when the receiver is 5 meters away from the transmitter, we observed that the throughput for all the agents including the Minstrel algorithm are zero for 5 000 and 10 000 packets per second arrival rate even though the energy consumption is higher than that in the test environment with only one jammer. However, the throughput gradually increases starting from  15 000 packets per second arrival rate: the throughput-critical agent has the highest throughput at about 600 Mbps, the agent with 0.5 $\lambda$ value increases to roughly 500 Mbps, the Minstrel algorithm has a maximum of 300 Mbps, and the energy-critical agent maximizes its throughput up to 200 Mbps.

Fig. \ref{fig:test10m2} is the result when the distance is extended to 10 meters. We observe a similar trend as the case with 5 meters distance but the Minstrel algorithm has much less throughput. Also, the energy-optimizing agent has almost zero throughput until the arrival rate reaches 40 000 packets per second . In terms of energy consumption, the agents and the Minstrel algorithm consume even higher energy than when there is only one jamming node. However, the energy expanded has a similar trend: less energy is consumed by the energy-optimizing agent, and higher energy is consumed by the throughput-optimizing agent.

Fig. \ref{fig:test20m2} shows the result when the distance is set to 20 meters with two jamming nodes, and we observe similar trend: the throughput-optimizing agent maintains roughly the same throughput as the scenarios with 5 and 10 meters distances while consuming even higher energy. The Minstrel algorithm's throughput is reduced to roughly 200 Mbps. The energy-optimizing agent's throughput gets drastically reduced to 50 Mbps, but it has less energy consumption than other agents in the same scenario.

\section{CONCLUSIONS AND FUTURE DIRECTION}

In this paper, we use RL to control both the transmission power and rate of an 802.11ac device under the impact of a jammer. The goal is to optimize a joint reward function consisting of throughput and energy consumption. With this reward function, users can define their priority to achieve either high throughput, longer battery life, or a balance of the two. By including several factors in the state of the environment, we are able to incorporate random factors such as interference and wireless channel fading into our system model. Extensive simulation is done using NS-3, and we have compared the optimal policies with the popular Minstrel algorithm. When we prioritize throughput, we are able to obtain much higher throughput than Minstrel. When the transmission distance is slightly increased, our approach maintains roughly the same performance in terms of throughput, but Minstrel suffers from significant decreased performance. For the throughput and energy-aware agent, which learns to optimize both the throughput and the energy consumed, we can achieve a higher throughput than Minstrel while consuming a slightly higher amount of energy. Finally, when the energy is prioritized in the reward function, our approach consumes less energy than Minstrel, but at the cost of reduced throughput.

We use tabular RL in the paper. One downside of it is the space complexity of the table. The agent may also not be able to take the optimal action if all the state-action pairs are not visited. Future direction will focus on introducing a Neural Network (NN) architecture which aids training the underlying model without having to traverse the whole state space. With the neural network architecture, we will not need to discretize the environment variables such as the packet queue length, the back-off slots, and the battery level. The agent will be able to compute the optimal policies for an unvisited state space.

\bibliographystyle{unsrtnat}
\bibliography{references}  






\end{document}